\documentstyle[11pt,aasms4]{article}

\begin{document}

\newcommand {\etal}{{\it et al.}}
\newcommand {\HI}{H~{\small I}}
\newcommand {\HII}{H~{\small II}}
\newcommand {\HA}{H$\alpha$}
\newcommand {\NII}{[N~{\small II}]}
\newcommand {\HN}{H$\alpha +$[N~{\small II}]}
\newcommand {\EW}{\HA\ EW}
\newcommand {\EA}{EW(\HA )}
\newcommand {\degree}{$^{\circ}$}
\newcommand {\Msolar}{$M_{\odot}$}
\newcommand {\Lsolar}{$L_{\odot}$}
\newcommand {\sfindex}{log($L_{{\rm FIR}}/L_{B}$)}
\newcommand {\fFIRfB}{$\log(f_{FIR}/f_{B})$}
\newcommand {\MHILB}{$M_{\rm H I}/L_{B}$}
\newcommand {\B}{$B^{0}_{T}$}
\newcommand {\BV}{$(B-V)^{0}_{T}$}
\newcommand {\LFIR}{$L_{{\rm FIR}}$}
\newcommand {\LB}{$L_{B}$}
\newcommand {\LHA}{$L_{{\rm H\alpha }}$}
\newcommand {\fHA}{$f_{{\rm H\alpha }}$}
\newcommand {\FIRHA}{log(\LFIR /\LHA )}
\newcommand {\rer}{$r_{e,R}$}
\newcommand {\rea}{$r_{e,{\rm H\alpha}}$}

\title{
Photometric Observations of 
Star Formation Activity in 
Early Type Spirals
}

\author{
Tadashi Usui and 
Mamoru Sait\={o}
}
\affil{
Department of Astronomy,
Faculty of Science, 
Kyoto University
\\
Sakyo-ku, Kyoto 606-8502, Japan
\\
Electronic mail:
usui, saitom@kusastro.kyoto-u.ac.jp
}

\author{
Akihiko Tomita
}
\affil{
Department of Earth and Astronomical Sciences,
Faculty of Education,
\\ Wakayama University, Wakayama 640-8510, Japan
\\
Electronic mail: atomita@center.wakayama-u.ac.jp
}

\begin{abstract}

We observationally study the current star formation activities of early type
spiral galaxies.
We construct a complete sample of 15 early type spirals having far-infrared 
(FIR) to optical $B$ band luminosity ratios, \sfindex , larger than the 
average of the type, and make their CCD imaging of 
the $R$ and \HA\ bands.
The equivalent widths of \HA\ emission increase with increasing 
\LFIR /\LB , indicating that  \sfindex\ can be an 
indicator of star formation for such early type spirals with star formation 
activities higher than the average.
For all of the observed early type spirals, the extended \HII\ 
regions exist at the central regions with some asymmetric features.
\HA\ emission is more concentrated to the galactic center than the 
$R$ band light, and the degree of the concentration increases with the star 
formation activity.
We also analyze the relation between the star formation activities and 
the existence of companion galaxies in the sample galaxies and other  
bright early type spirals.
No correlation is found and this suggests that the interaction is not 
responsible for all of the star formation activities of early type spirals.

\end{abstract}


\section{Introduction}

\cite{K94} (1994) extended the study of star formation activities of spiral 
galaxies by \cite{K83} (1983), and 
showed that the ratio of the current star formation 
rate (SFR) to the average past SFR increases as the morphological type 
becomes later, from 0.01 in Sa to 1 in Sc. 
They used the integrated \HA\ equivalent width (EW) as an indicator of star 
formation activity.
Their samples were about 200 galaxies, 
in which galaxies having active galactic nuclei (AGNs) or interacting 
companions were not included.

 On the other hand, far-infrared (FIR) luminosity of spiral galaxies 
is considered to be another indicator of the current SFR.
A luminosity function of 60 $\mu$m 
emission obtained by the {\it IRAS} 
is similar for galaxies from Sa to Scd (\cite{DY91} 1991).
FIR to optical $B$ band luminosity ratio, \sfindex , has been used as 
an indicator of the present star formation activity of spiral galaxies 
by many researchers ( e.g., \cite{deJ} 1984, \cite{Both} 1989, 
\cite{tomita96} 1996, \cite{DH} 1997).
\cite{tomita96} (1996) constructed a histogram of \sfindex\ 
for each morphological type, and showed that the distribution is 
rather flat from 
$-$1.5 to 0.5 in early type spirals (Sa-Sab), while it is much 
concentrated within 1 dex around $-$0.5 in late type spirals (Sc).
These suggest that there are many early type spirals with high star formation 
activities, in contrast with the result of \cite{K94} (1994).

 There is, however, a controversy over the origin of FIR emission of 
spiral galaxies. 
FIR emission generally consists of 
at least two components: a warm component from dust in star forming region and 
a cool component from diffuse interstellar dust, heated by the general 
interstellar radiation field (e.g., \cite{He86} 1986).
\cite{DY90} (1990) investigated the relation between 
FIR and \HA\ luminosity and showed that high-mass stars are responsible 
for the both luminosities. 
The morphological dependence of the ratio of FIR to \HA\ luminosity 
was investigated by \cite{ST92} (1992).
Their  ratios systematically decrease from early to late type 
spirals, and they interpreted this trend as the decreasing contribution 
of cirrus component, from about 86\% for Sa galaxies to about 3\% for Sdm 
galaxies.
\cite{Xu90} (1990) analyzed intensities of UV (at 2000 \AA ) 
and FIR emission 
for about 40 spirals, most of which are late type, and concluded that 
the cool component is dominantly due to the 
non-ionizing UV radiation from intermediate massive stars ($\sim$ 5 \Msolar ) 
with lifetimes of the order of $10^8$ yr (see also \cite{Walt96} 1996).
Moreover, there is a tight and universal correlation between the integrated 
FIR and radio continuum emission, most of which is non-thermal, 
for various types of galaxies from Sa to Sm (\cite{Wnd} 1987).
The radio continuum is generated by the synchrotron 
radiation of cosmic ray electron, and cosmic rays are accelerated by 
supernova explosions, whose progenitor is also responsible for FIR emission 
(\cite{Lis} 1996).
This correlation further supports the notion that FIR emission is a star 
formation indicator as well as \HA\ emission.

 The sample of early type spirals of \cite{K83} (1983) and \cite{K92} 
(1992), on which \cite{ST92} (1992) and \cite{K94} (1994) were based, 
is small in number and may be biased to the galaxies with low values 
of \sfindex . 
\cite{Cald} (1991) studied star formation activities of Sa galaxies,
selected from the Revised Shapley-Ames Catalog (\cite{RSA} 1987). 
But most of their galaxies are weak in FIR emission, i.e., 
\sfindex\ $< -$1.0, and half 
of their sample are classified as earlier than Sa in 
the Third Reference Catalog of Bright Galaxies (RC3; \cite{deV91} 1991).
Recently \cite{yng96} (1996) presented the \HA\ emission line flux of 
spiral galaxies extensively, 
but the sample is still small in number for early type spirals with 
high values of \sfindex .

In this study we make CCD imaging of the $R$ and \HA\ bands for 15 early type 
spirals with \sfindex\ $\ge -$0.5 and investigate the properties of 
FIR emission as an indicator of star formation activity by comparing 
with \HA\ emission.
\HA\ emission originates exclusively from newly born massive stars, aside 
from AGN, and connects directly with the present 
star formation activities, though the effect of the 
extinction remains uncertain.
Moreover the \HA\ images provide some insights about the distribution of 
star forming regions. 
We describe the sample selection, observations and data reduction in $\S$~2.
In $\S$~3, after comments on individual galaxies, we show that \sfindex\ 
can be a star formation indicator comparable to \HA\ luminosity for 
early type spirals with higher star formation activities.
We also show the trend of the central concentration of the star forming 
regions in the observed galaxies.
In $\S$~4 we discuss the difference of star formation indicators between 
FIR and \HA\ emission, and estimate a general distribution of \EW\ of early 
type spirals.
We also analyze the effect of galaxy-galaxy interactions on the star formation 
activities in early type spirals.
A summary is given in $\S$~5. 

\section{Data}
\label{Dat}

\subsection{Galaxy Sample}
\label{Samp}
In this paper we refer early type spirals as galaxies with morphological 
types Sa and Sab, or $0.5 \le T < 2.5 $ for the index of 
the Hubble sequence given in RC3. 

We select early type spirals having \B\ and \BV\ in RC3
and \sfindex\ $\ge -$0.5.
The range of \sfindex\ adopted here represents that the selected galaxies have 
\sfindex\ higher than the average of early type spirals.
The values of \sfindex\ are computed  using \B\ in RC3 and 60 $\mu $m and 
100 $\mu $m flux densities of {\it IRAS} data in the same manner as that of  
\cite{tomita96} (1996). 
The FIR luminosity, \LFIR , in solar units is given by \cite{Lons} (1985) as
\begin{equation}   
\mbox{log\LFIR }=\mbox{log}(2.58f_{60}+f_{100})+2\mbox{log}D+5.595,
\end{equation}   
where $D$ is the distance in Mpc, assuming $H_0$ = 75~km~s$^{-1}$~Mpc$^{-1}$, 
{\it f}$_{60}$ and {\it f}$_{100}$ are the flux densities 
at 60 $\mu $m and 100 $\mu $m in Jy.
Following \cite{Soif} (1987), the $B$ band luminosity, \LB , in solar units 
is given as
\begin{equation}   
\mbox{log\LB }=-0.4B_T^0+2\mbox{log}D+11.968.
\end{equation}   
The additional criteria of our sample are as follows: 
they are 
(1) nearby galaxies with recession velocities 
1000~km~s$^{-1}~\le~cz~\le~2600~$km~s$^{-1}$,  
(2) at the declination range $-15$\degree\ $\le \delta \le 70$\degree\ , 
(3) with optical diametars $\le $~4.$'$0 to secure sufficient sky area in 
 each CCD image,
(4) rather face-on with inclination  $\le 60$\degree\ , which suffers 
less extinction in \HA\ luminosity (\cite{yng96} 1996),  and 
(5) without Seyfert activity. 
The sample selection is not related to the presense of a bar nor a sign of
interaction, because our main purpose is to obtain the relation between 
\HA\ and FIR emission of early type spirals as a whole.

The complete sample consists of 15 galaxies, which are given in 
Table~\ref{tbl-1} along with the properties.
\B\ of the sample galaxies ranges from 11.5 to 13.6.
The range of optical diametars is from 10~kpc to 20~kpc, except for two 
small galaxies and one large galaxy.
Seven sample galaxies were previously identified as Markarian or Kiso 
Ultraviolet Excess Galaxies (KUGs), as shown in the 15th column of 
Table~\ref{tbl-1}.
They are UV-bright and suggest higher star formaion activities.
Nuclear specta have been known to be the \HII\ region type for 8 our sample 
galaxies in the literature (column 13 of Table~\ref{tbl-1}).

We also observed other 6 spiral galaxies regardless of morphological type for 
a comparison, \EW s of which had been measured by other reserchers.
These galaxies are listed in Table~\ref{tbl-2} with the properties.
NGC~681 is one of our sample galaxies, and the \EW\ of the galaxy was 
measured by \cite{K83} (1983).
So we use this galaxy as a comparison galaxy.


\subsection{Observations and Data Reduction}
\label{Obs}
 The observations were made using a CCD camera (Thomson 7882, 576 $\times384 $ 
pixels) attached to the Cassegrain focus of the 0.6 m reflector 
(focal length = 4.8 m) at the Ouda Observatory, Kyoto University 
(\cite{ohtani92} 1992) between 1996 December and 1998 March.
The scale was 1.$''$0 per pixel and the field of view was 10$'$ 
$\times$ 6$'$.
Seeing during the run was around 5$''$. 
To measure \HN\ emission line intensities, we observed each galaxy using 
two filters, i.e., a narrow band filter ($\lambda_{c} = 6610$~\AA\ , 
$\Delta \lambda = 80$~\AA\ ), which covers \HN\ emission for 
galaxies with 
recession velocities between 1000~km~s$^{-1}$ and 3500~km~s$^{-1}$,
and a broad $R$ band filter with a band width of $\sim $ 1000~\AA , 
which gives the continuum intensity around the emission lines.
The integration time for each frame was 2 to 4 min, and the total integration
time of each object was 1 to 2 h in the 
narrow band and 10 to 20 min in the $R$ band. 

Standard CCD reductions were performed using 
IRAF\footnote{IRAF is the software developed in National Optical Astronomy 
Observatories.}.
The images were bias-subtracted and flat-fielded using a frame of 
twilight sky.
Then the images were sky-subtracted, aligned and scaled using foreground 
stars, and median-combined to make the final image of each galaxy in 
the both narrow and $R$ band.
To determine the continuum level, the $R$ band image was scaled by the 
ratio of the count of the foreground stars of the $R$ band image to that 
of the narrow band image.
The scaled $R$ band image was subtracted from the narrow band image, and the 
pure \HN\ emission line image was yielded.
We determined the continuum level with an error of around 5\%. 
The uncertainty depended on mainly the number of forground stars in each image.

We obtained the EW of \HN\ emission lines as the ratio of 
the \HN\ emission line flux 
to the continuum flux in the aperture which was taken from $D_{25}$ in RC3. 
\NII /\HA\ ratio in the integrated light of each galaxy is considered to 
be similar from galaxy to galaxy (\cite{KK83} 1983).
So we did not correct for the \NII\ emission, and hereafter we refer the 
EW of \HN\ emission lines as \EW\ or \EA .
The second column of Table~\ref{tbl-3} shows the \EW s with the uncertainties 
for the sample galaxies as well as the comparison galaxies.
The uncertainties were derived in two ways; one is from the statistics of the 
sky level and the scaling factor to determine the continuum level, and 
another is from the internal disagreement of different nights' observations 
of the galaxy. 
In most cases the two uncertainties were comparable; 
if they were significantly different, the larger value is listed in Table~\ref{tbl-3}.


We estimated \HN\ emission line flux, which hereafter we refer as 
\fHA , in the following manner.
The $R$ magnitude has not given for most of the sample galaxies, but 
all of the observed galaxies have $B_T$ and $V_T$.
First, we constructed the $(B-V)_T-(V-R)_T$ diagram by combining 
the data of $(B-V)_T$ in RC3 and $(V-R)_T$ in \cite{Bt95} (1995) for 
mostly normal galaxies.
Then we estimated the $R$ magnitude from B$_T$ and V$_T$, using the equation: 
$R_T = V_T - (V-R)_T = V_T - 0.37 (B-V)_T - 0.23 \pm 0.05$.
We converted the $R$ magnitude to the flux density, and assumed it to 
be the continuum flux density around the \HA\ line. 
\fHA\ was computed from \EW\ and the flux density.
\HN\ luminosity in solar units, which hereafter we refer as  \LHA , 
was also estimated after the correction for the Galactic extinction.
The Galactic extinction in $B$ band, $A_g$, in RC3 was converted to the 
extinction at \HA\ line, assuming $A$(\HA )/$A_g$~=~0.61.
We did not correct for the internal extinction nor \NII\ emission.
The estimated \fHA\ and \LHA\ are given in the third and fourth columns of 
Table~\ref{tbl-3}, respectively.
\LHA\ of the sample galaxies are of the order of $10^7$~\Lsolar .
The data of \cite{Ho97} (1997) showed that \LHA\ of LINERs are less than 
2~$\times$~10$^{6}$~\Lsolar\ for all of the observed 28, except one, 
early type spirals. 
So in our sample galaxies the contribution of LINERs, if any, is probably 
negligible in the \LHA .

We also measured half-light radii of the $R$ band continuum and \HA\ 
emission, \rer\ 
and \rea , respectively, to gauge the distribution of light within a 
galaxy.
The half-light radii were defined such the radius that the circular 
aperture of \rer\ (\rea ) encompasses a half of the total flux of 
$R$ (\HA ) band light of the galaxy.
The measurements on the different nights' imagings of each galaxy agree 
quite well.
We calculated the ratio of \rea\ to $r_{25}$ and the 
ratio of \rea\ to \rer , where $r_{25}$ is the optical radius of the 
galaxy and equal to a half of $D_{25}$. 
\rea /$r_{25}$ indicates the degree of the concentration of 
\HA\ emission relative to the optical size of the galaxy, and 
\rea /\rer\ indicates the degree of the concentration of 
\HA\ emission relative to the continuum light.
We tabulate \rea\ in kpc, \rea /$r_{25}$ , and \rea /\rer\ in the last 
three columns of Table~\ref{tbl-3}. 

\section{Results}
\subsection{Comparison with Previous Studies}
\label{Comp}
We observed 7 galaxies listed in Table~\ref{tbl-2} with known \EW s.
Figure~\ref{f1} shows a comparison of \EW s measured by us with those 
measured by \cite{K83} (1983) and \cite{Rom90} (1990).
Their measurements were made by using large aperture (1$'$ to 7$'$) 
photometry.
The \EW s of \cite{K83} (1983) were corrected by a factor 1.16 following 
by \cite{K94} (1994).
The mean error is : 
\begin{equation}   
\Delta[\mbox{\EA\ (this study)}-\mbox{\EA\ (other studies)}] = 
2.6 \pm 6.7~\mbox{\AA }. 
\end{equation}    
Our values of \EW s thus agree with the values obtained by others 
for \EA\ $>$ 20~\AA  ; most of the sample galaxies have \EW s in this range, 
as shown in Table~\ref{tbl-3}.
NGC~681 has relatively weak \EW s and the difference between \cite{K83}'s 
(1983) and ours is the largest.
This galaxy is one of our sample, and the physical values derived from 
the two \EW s shall be shown hereafter.

We also compared our estimated \fHA\ with the values of \cite{K83} (1983) 
and \cite{Rom90} (1990), and 
the ratio is :
\begin{equation}   
\mbox{log}\left[\frac{ \mbox{\fHA\ (this study)} }{ \mbox{\fHA\ (other studies)} }\right] = -0.03 \pm 0.09. 
\end{equation}   
These errors are small enough for the following discussion.


\subsection{Description of Individual Galaxies}
\label{Indi}
 The images of 15 sample galaxies observed by us are shown in Figure~\ref{f2}.
For each galaxy, the $R$ band image is left and the \HA\ image is right.
North is at the top and east is at the left.
The intensity is scaled logarithmically. 
Though the resolution of our images is around 5$''$, it is apparent that 
there are much variety in \HA\ properties.
We describe the images of individual galaxies briefly.


NGC~681. The inclination of NGC~681 estimated is small and close to the 
critical value of our sampling.
The \HA\  emission is strong in the central region  and extended along 
the disk. 
A dust lane is seen on the north of the center.
The bright object at the north-western edge of this galaxy is a 
foreground star.

NGC~1022. In the Carnegie Atlas of Galaxies (\cite{San} 1994, 
the Carnegie Atlas), thin dust 
filaments are seen within the ring, but they cannot be seen in our $R$ band 
image because of the poor resolution.
The \HA\ emission is concentrated to the central region.

NGC~2782. This galaxy is sometimes classified as a Seyfert galaxy, but 
 \cite{boer} (1992) showed that the nuclear spectrum is typical of nuclear 
starbursts and the high excitation gas due to shocks is on the 4$''$ to 8$''$ 
south of the center.
In the Carnegie Atlas, low surface brightness components extend 
far beyond $D_{25}$, and fragments of dust lanes in spiral arms exist 
through the central bulge.
\cite{Smith} (1994) considered this galaxy is merging with a dwarf galaxy.
\HA\ emission is strong in the central region and also seen on the 
north-western arc.
The southern component which corresponds to the high excitation gas is weak 
in \HA\ emission compared with the center.

NGC~2993. This galaxy is strongly interacting with NGC~2992, which is at the 
3$'$ north-west of the galaxy.
Tidal tails are seen in our original image and the Carnegie Atlas.
Star formation occurs in the central region, and diffuse \HA\ emission 
surrounds it.

NGC~3442. This galaxy is small and less luminous.
No feature is seen in the $R$ band image.
In the \HA\ image two discrete \HII\ regions are seen across the nucleus. 

NGC~3504. The \HA\ emission is strong in the galactic center and also 
on the inner ring.
The brightest part on the ring coincides with the southern end of the bar.
There is a companion galaxy, NGC~3512, at 12$'$ east of this galaxy.
This galaxy is extensively studied by \cite{Ke93} (1993).

NGC~3611. This galaxy  has a companion galaxy UGC 6306 at the south with 
a separation of 3$'$.
The \HA\ emission is concentrated to the central region, and the 
emission region extends toward south.

NGC~3729. Both the $R$ and \HA\ images of this galaxy resemble NGC~3504, 
though the star formation activity is weaker than that galaxy.
The bright object at the southern edge of this galaxy is a foreground star.

NGC~4045. In the $R$ band image the outer envelope is rather oval, 
but the inner ring is almost circular.
The inner ring is also traceable in \HA\ emission.
The brightest \HII\ region is at the galactic center. 
NGC~4045A ($cz$ = 5329 km~s$^{-1}$ ), which appears on the southern edge 
of this image, does not have a physical relation with the galaxy.

NGC~4369. The $R$ band image is featureless.
Discrete two \HII\ regions are resolved around the galactic center, 
but the galactic center itself is faint in \HA\ emission.

NGC~4384. Two major \HII\ regions are seen symmetrically with respect to 
the galactic center, and diffuse \HA\ emission surrounds them. 
In the {\it HST} image, spiral structure reaches down to the center, and 
several bright sources, likely \HII\ regions, can be seen (\cite{Car} 1997).

NGC~5534. This galaxy may be in the process of merger (see the Carnegie Atlas).
The strong \HA\ emission is in the central region, and there are emission 
regions at the western side of the galaxy.
The companion galaxy is located at the eastern side and also shows \HA\ 
emission.

NGC~5691. In the $R$ band image, there is a dust lane at the north-west of the 
galactic center.
The \HII\ regions are seen at the central region.

NGC~5915. This galaxy is interacting with NGC~5916 and NGC~5916A, which are 
located at 5$'$ south and 5$'$ west of the galaxy, respectively.
The \HA\ emission is seen on the entire galaxy and the brightest \HII\ region 
is on the southern arm, which is on the side of NGC~5916.

NGC~7625. This galaxy has a tube-like dust lane in the Carnegie Atlas, 
which cannot be seen in our $R$ band image.
Patchy \HII\ regions and more diffuse emission are seen in the \HA\ image.
This Arp galaxy occurs intense star formation, but does not have any companion 
galaxy.
\cite{Li93} (1993) observed and studied this galaxy extensively.

The early type spirals observed in the present study have the 
extended \HII\ regions at the central regions with some asymmetric 
features.
The intensities at faint extended \HA\ emission are well above the errors 
estimated from the uncertainties of continuum subtraction and scaling 
factor between \HA\ and $R$ band image, and these features are real.
For some of barred or ringed galaxies, e.g., NGC~3504, NGC~4045, the 
star formation also takes place on the ends of a bar or a ring.
Nevertheless it is at the central region that the star formation takes place 
most intensely.
The degree of the concentration of the \HII\ regions to the galactic 
center is discussed 
quantitatively in $\S$~\ref{Dist}. 

\subsection{Correlation between \EA\ with $L_{{\rm FIR}}/L_B$ }
\label{EW}
 In Figure~\ref{f3} we plot the log[\EA ] versus \sfindex\ relation for 15 
sample early type spirals by filled circles as well as 16 early type 
spirals of \cite{K83} (1983, 1992) by open circles.
For their galaxies having two data, we adopted the mean \EW .
We show two \EW s of NGC~681 obtained by us and \cite{K83} (1983), 
and connected each data point by a vertical line.
The mean of the two values is also shown by a cross.
In Figure~\ref{f3} there is a trend that log[\EA ] increases with increasing \sfindex .
A least-squares fit yields 
\begin{equation}   
\mbox{log[\EA ]}=0.82\mbox{\sfindex }+1.48,
\end{equation}   
and the correlation coefficient is 0.79, provided that we excluded 
galaxies with \EA\ $\le$ 3~\AA\ from the fit because of the large errors 
of \EW\ in logarithmic scale.
Thus \sfindex\ can be a star formation indicator for early type spirals 
with higher star formation activities.
We do not extrapolate the correlation to less active galaxies because of 
the presence of two sources of the FIR emission, i.e., cirrus and weak 
\HII\ regions, and difficulty in measurements of weak \EW . 
It can only be said that galaxies with \sfindex\ $< -0.5$ have \EW s less 
than 10~\AA .
In $\S$~\ref{diff} we shall interpret the non-linearity and relatively 
large scatter of the correlation between \EW\ and \LFIR /\LB .


\subsection{Star Forming Region }
\label{Dist}
We regard \rea\ listed in Table~\ref{tbl-3} as a typical radius of star 
forming region of each galaxy. 
For the sample galaxies the mean of \rea /r$_{25}$ is 0.15, and 
\rea /\rer\ is less than unity for most of galaxies.
The former means that a half of star formation occurs within the inner 
15\% of the optical radius.
The latter means that \HA\ emission is more concentrated to the central 
region than the $R$ band light.
\cite{RD94} (1994) observed 34 spiral galaxies, most of which are late type, 
and showed that the scale length of \HA\ emission is much longer than the $V$ 
and $I$ scale length in the outer disk, indicating that \HA\ emission is 
{\it less} concentrated to the galactic center than the broad band light 
in late type spirals.

In Figure~\ref{f4} we plot \rea /\rer\ as a function of \sfindex\ for 
our sample galaxies (filled circles).
One comparison galaxy, NGC~7217, is classified 
as an early type spiral but we did not include this galaxy in our sample 
because of the low value of \sfindex .
We plot the data of NGC~7217 by a open circle for a comparison. 
In the diagram the galaxies with vertical bar are barred galaxies (SB), and 
ones with horizontal bar are galaxies with companions.
There is a trend that the degree of the concentration increases as the 
star formation activity is higher.
The trend seems to be not related to the presence of companion nor bar 
structure. 
\cite{LH96} (1996) reported that \rea /\rer\ is weakly correlated 
with the FIR color but not correlated with the ratio of IR-to-$B$ band 
luminosity for more actively star forming galaxies than ours.
But the high extinction and the edge-on nature of their sample may blur 
the correlation if it exists.


\section{Discussion}
\subsection{Difference of the Star Formation Indicators in \HA\ and FIR}
\label{diff}
Though \LFIR /\LB\ correlates with \EW\ (see Figure~\ref{f3}), there are some 
differences between \HA\ and FIR emission as the star formation 
indicators.
Figure~\ref{f5} shows log(\LFIR /\LHA ) as a function of \sfindex\ for 
early type spirals with \sfindex\ $\ge -0.5$ of our sample (filled circles) 
and the sample of \cite{K83} (1983, 1992) (open circles), which have \EA\ 
$\ge 10$ \AA , except for two Kennicutt's objects including NGC~681 
(see Figure~\ref{f3}).
For the objects of \cite{K92} (1992), 
we estimated \fHA\ from \EW\ by the same manner as described in $\S$~\ref{Obs}.
NGC~4750 has no (B$-$V)$_T$ data in RC3, and we omitted the galaxy from 
their sample. 
\LHA\ was corrected for the Galactic extinction and not corrected for the 
internal extinction nor \NII\ emission.
For the plotted galaxies, there is a trend that \FIRHA\ increases with 
\sfindex\ with a correlation coefficient of 0.60; as the star formation 
activity is higher, \HA\ emission tends to become weaker compared with 
FIR emission.
This trend corresponds to the fact that the slope of Figure~\ref{f3} 
is less than unity, as shown in equation (5).
The feature originates due to the heavier 
obscuration for younger stellar objects.
\cite{YSO} (1989) found that in the Galaxy O stars are obscured in the 
parent molecular clouds during a time of 10\% to 20\% of the life; in the 
process of star formation \LFIR\ grows faster than \LHA .
Parent interstellar matter removes from young massive stars with time, 
the intensities of FIR emission reduce, and the \HII\ regions become apparent.
So \LHA\ begins to grow later than \LFIR\ and is increasing even at the 
phase that \sfindex\ is decreasing.

On the other hand, Figure~\ref{f4} suggests that in our sample galaxies 
the more active star formation occurs in the more concentrated region.
This concentration may also affect the value of \FIRHA . 
In Figure~\ref{f6} we plot \FIRHA\ and \rea /\rer\ for our sample early type 
spirals (filled circles) and a comparison early type spiral (open circle).
Although the correlation between the two parameters is marginal, 
there is a weak trend that the higher central concentration (i.e., lower \rea 
/\rer ) leads to the higher \FIRHA .
The extinction for \HA\ emission possibly becomes higher towards the galactic 
centers.

The value of \FIRHA\ thus depends on both the phase of the star formation 
and the extinction for early type spirals with higher star formation 
activities, and the both effects yield the 
non-linearity and the relatively large scatter of the correlation 
between \LFIR /\LB\ and \EW\ in Figure~\ref{f3}.
                    

\subsection{Distribution of \EW\ of Early Type Spirals}
\cite{tomita96} (1996) made a universal histogram of \sfindex\ for each 
morphological type.
We divide \EW\ into three ranges: 0~-~10 \AA , 10~-~20 \AA , and 
over 20 \AA .
For early type spirals the range of \EA\ = 10~-~20 \AA\ corresponds to 
\sfindex\ $\simeq -$0.47 to 0, and the range of \EA\ $>$ 20 \AA\ 
corresponds to \sfindex\ $>$ 0, from equation (5).
The universal histogram of \sfindex\ by \cite{tomita96} (1996) indicates 
that the relative frequency of the two ranges of \EW\ are, respectively, 
17\% and 29\% for early type spirals.
Because of the scatter of the relation between \EW\ and \LFIR /\LB , 
the conversion of \EW\ to \sfindex\  may not be so straightforward, and this 
estimation is rather crude.
These values suggest, however, that the star formation activity of early 
type spirals is low [\EA\ $\le$ 10 \AA ] for half or somewhat more 
galaxies, and the star formation activity is modest or strong 
[\EA\ $\ge$ 20 \AA ] for about 30\% of early type spirals.
The star formation activity of early type spirals as a whole is lower than 
that of late type spirals, \EW s of which are around 30 \AA\ 
(\cite{K83} 1983), 
but the difference is not so large as \cite{K94} (1994) showed.
The sample which they used is biased to the galaxies with 
\sfindex $\le -$0.5 (see Figure~\ref{f3}), while the average of \sfindex\ is 
$-0.5$ for early type spirals. 

\subsection{Relation between the Star Formation Activities and 
Galaxy Interactions }
\label{Inter}

Galaxy-galaxy interactions are considered to be an important triggering 
mechanism for starburst galaxies (see a review by \cite{Barnes} 1992).
Both the integrated \HA\ emission and the FIR emission are systematically 
enhanced in the strongly interacting systems like galaxies in the 
Atlas of Peculiar Galaxies (\cite{Arp} 1966), 
but enhancements are subtle for nearby pair galaxies (\cite{K87}, 1987).
We examine the relation between interactions and star formation activities 
in nearby early type spirals by the same method as that of 
\cite{Dahari} (1985).
We assume that the tidal force is proportional to $M_{C}M_{P}^{-1}r^{-3}$, 
where $M_P$ and $M_C$ are the masses of the parent galaxy and its companion, 
respectively, and $r$ is the spatial separation between the galaxies.
We use the $B$ band luminosity instead of the mass of the galaxy.
We also assume that the amount of energy exchanged in the encounter 
is proportional to $|\Delta V|^{-2}$, where $\Delta V$ is the relative 
velocity between the pair members.
We can practically know only the projected spatial distance, $S$, and the 
velocity difference along the line of sight, $|c\Delta z|$.
So we define the interaction parameter (IP) by
\begin{equation}   
\mbox{IP}=\mbox{log}L_C-\mbox{log}L_P-3\mbox{log}(Sz)-2\mbox{log}|c\Delta z|.
\end{equation}   
 Here $L_C$ and $L_P$ are the $B$ band luminosity of the companion  
and the parent galaxy, respectively.
If $L_C$ is unknown, we assumed the mass is proportional to $D_{25}^{1.6}$, 
where $D_{25}$ is the optical diameter in RC3.
In this case IP is defined by 
\begin{equation}   
\mbox{IP}=1.6\mbox{log}D_C-1.6\mbox{log}D_P-3\mbox{log}(Sz)-2\mbox{log}|c\Delta z|,
\end{equation}   
where $D_C$ and $D_P$ are $D_{25}$ given in RC3 of the 
companion and the parent galaxy, respectively.
In this case, IP is identical to Q$'$ in \cite{Dahari} (1985).
IPs derived by above equations are the upper limits in the sense that we adopt 
the minimum values of $\Delta V$ and $r$. 

We searched for companion galaxies within the radius of 10$D_P$  
and redshift difference less than 1000 km~s$^{-1}$ in RC3, 
and calculated IP for each companion galaxy.
If there were two or more companion galaxies in the searched space, 
we adopted the largest value of IP among IPs calculated for each 
host-companion galaxy pair.
IPs were derived for 87 early type spirals with \B\ $\le$ 12~mag in RC3 and 
for our sample galaxies.
The distributions are shown as a function 
of \sfindex\ in Figure~7(a) and 7(b), respectively.
In these diagrams we adopted IP = $-$5 for galaxies without any companion 
galaxy.
The galaxies with cross in both diagrams of Figure~\ref{f7} are 
classified as peculiar in RC3.
The peculiar galaxies are considered to have high 
IP and high \sfindex ; a peculiarity connects with both 
the presense of companion galaxies and enhanced star formation.
In Figure~\ref{f7}, however, the relation between IP and \sfindex\ is random 
on the whole, and there are star forming early type spirals with neither 
companion nor peculiarity.
There is no companion galaxy for more than half of our sample galaxies 
as shown in Figure~7(b).
We conclude that interactions are not responsible for all of the star 
formation activities of early type spirals.

On the other hand, bar structures may enhance star formation activities in 
early type spirals to some extent.
\cite{Huang} (1996) showed that \sfindex\ of barred galaxies (SB) is 
systematically higher than that of un- and weakly-barred galaxies (SA and 
SAB) only for early type spirals (S0/a - Sbc).
But the relation between \rea /\rer\ and \sfindex\ is similar regardless 
of the presence of bars or companion galaxies (see Figure~\ref{f4}). 
Further observations are needed to fully understand the star formation 
mechanism(s) of early type spirals.


\section{Summary}
\cite{K94} (1994) used \EW\ as an indicator of the present star formation 
activities and showed that star formation activities of early type spirals 
are lower than those of late type spirals by a factor of 100. 
On the other hand, \cite{tomita96} (1996) used \sfindex\ as an indicator 
of star formation activities, constructed the histogram of \sfindex\ of 
each morphological type, and suggested that there are many early type 
spirals with high star formation activities.
In order to solve the inconsistency, we made \HA\ imaging observation of 
the complete sample of 15 early type spirals 
with higher \sfindex\ than the average of the type, and have obtained the 
following conclusions.

1.~We have compared \sfindex\ and log[\EA ] of early type spirals for 
30 early type spirals consisting of our sample and \cite{K83}'s (1983, 1992) 
sample, and obtained the correlation that \LFIR /\LB\ increases with 
increasing \EW . 
\sfindex\ can be a star formation indicator for early type spirals with 
star formation activities higher than the average.

2.~For the early type spirals observed by us, the extended \HII\ 
regions exist at the central regions with some asymmetric features.
Additional \HA\ emission is seen on the ends of a bar or a ring for 
some barred or ringed galaxies. 
Even in such cases, it is at the central region that the star formation 
takes place most intensely.
\HA\ emission is more concentrated to the galactic center than the 
$R$ band light, and the degree of the concentration increases as the star 
formation activity is higher.

3.~We have found that \FIRHA\ tends to increase with increasing of both 
the activity and the central concentration of the star formation for 
galaxies with \sfindex\ $> -0.5$. 
These features are interpreted as the composite effects of 
the extinction and the phase of the star formation.
These effects yield the non-linearity and scatter in the 
correlation between \LFIR /\LB\ and \EW .

4.~Our results suggest that star formation activity is modest or strong 
[\EA\ $\ge $ 20 \AA ] for about 30\% of nearby early type spirals as a whole.

5.~Only about half of our sample galaxies have companion galaxies.
There is no correlation between star formation activities and the 
degree of the interaction for early type 
spirals brighter than 12~mag as well as our sample galaxies.
This implies that the galaxy interaction is not responsible for 
all of the star formation activities of early type spirals.

\acknowledgements

We would like to thank Yuuji Yamamoto, Yoshio Tomita, Taichi Kato and 
Tsuyoshi Ishigaki for their help with the observations.
This research has made use of the NASA/IPAC Extragalactic Database 
(NED), which is operated by the Jet Propulsion Laboratory, Caltech, under 
contact with the National Aeronautics and Space Administration.

\begin{table}
\dummytable\label{tbl-1}
\dummytable\label{tbl-2}
\dummytable\label{tbl-3}
\end{table}

\newpage

\figcaption[Usui.f1.ps]{Comparison of \EW s of this study with those of other 
studies for 7 galaxies.
The diagonal line is shown for a reference.
\label{f1}
}

\figcaption[Usui.f2a.ps,Usui.f2b.ps]{$R$ band (left) and 
continuum-subtracted \HN\ (right) images of the sample galaxies.
North is at the top and east is at the left in each image.
The images are trimmed to 3.3$'\times$3.3$'$, and the scale of 1$'$ is shown 
in the \HA\ image of NGC~681 and NGC~4045.
The intensity is scaled logarithmically.
\label{f2}
}

\figcaption[Usui.f3.ps]{ log[\EA ] as a function of  \sfindex\ for early type 
spirals observed by us (closed circles) and \cite{K83} (1983, 1992) 
(open circles).
\EW s of  NGC~681 observed by us and \cite{K83} (1983) are 
connected by a vertical line, and the mean of the two value is also 
shown with a cross.
The dashed line is the least-square fit for galaxies with \EA\ $\ge $ 3 \AA .
\label{f3}
}

\figcaption[Usui.f4.ps]{\rea /\rer\ as a function of \sfindex\ 
for our sample galaxies plus one galaxy, NGC 7217, from the comparison 
sample which is classified as an early type spiral. 
The galaxies with vertical bar are barred galaxies (SB), and 
ones with horizontal bar are galaxies with companions.
\label{f4}
}

\figcaption[Usui.f5.ps]{\FIRHA\ as a function of \sfindex\ for galaxies 
with \sfindex\ $\ge -$0.5.
The symbols are the same as in Fig.~3.
The dashed line is the least-square fit.
\label{f5}
}

\figcaption[Usui.f6.ps]{\FIRHA\ as a function of \rea /\rer .
The sample is the same as in Figure~\ref{f4}.
\label{f6}
}

\figcaption[Usui.f7a.ps,Usui.f7b.ps]{The interaction parameter, IP, as a 
function of \sfindex\ for (a) 87 early type spirals with 
\B\ $>$ 12 mag and (b) our sample galaxies.
The galaxies with cross are classified as peculiar in RC3.
\label{f7}
}

\clearpage
\begin{deluxetable}{ccclccccc}
\footnotesize
\tablecolumns{9}
\tablewidth{0pt}
\tablenum{2}
\tablecaption{Global Propeties of the Comparison Galaxies}
\tablehead{
\colhead{Galaxy}  & \colhead{$\alpha$(2000)}  &  \colhead{$\delta$(2000)}  &  
\colhead{Type}  &  \colhead{$B_T^0$}  &  \colhead{$V_{GSR}$~}  &  
\colhead{$D_{25}$} & \colhead{EW(H$\alpha$)} & \colhead{Ref.}  \\
\colhead{} &  \colhead{(h m s)} &  \colhead{~~($^{\circ}$ $'$ $''$)} & 
\colhead{} &\colhead{} &  \colhead{(km~s$^{-1}$)} &  \colhead{($'$)} & 
\colhead{(\AA )} & \colhead{} \\
\colhead{(1)} &  \colhead{(2)} &  \colhead{(3)} &  \colhead{(4)} &  
\colhead{(5)} &  \colhead{(6)} &  \colhead{(7)} &  \colhead{(8)} &  \colhead{(9)} }
\startdata
NGC   \phn681\tablenotemark{a}   & 014910.9 & $-$102540 &   SAB(s)ab sp & 12.50 & 1769 &  2.6 &  \phn4$\pm$2 &  1  \nl
NGC  3055  & 095517.4 & +041617 &  SAB(s)c & 12.18 & 1698 &  2.1 &  59$\pm$4 & 2 \nl
NGC  6207  & 164304.4 & +364959 &   SA(s)c & 11.57 & 1010 &  3.0 &  35$\pm$4 &  1  \nl
NGC  6574  & 181150.7 & +145850 &  SAB(rs)bc: & 11.79 & 2441 &  1.4 &  27$\pm$5  & 1 \nl
NGC  7217  & 220752.2 & +312135 &  (R)SA(r)ab & 10.53 & 1162 &  3.9 &   \phn6$\pm$2 & 1 \nl
NGC  7448  & 230003.7 & +155857 &  SA(rs)bc & 11.50 & 2370 &  2.7 &  40$\pm$4 & 1 \nl
NGC  7743  & 234421.6 & +095604 &  (R)SB(s)0+ & 12.16 & 1807 &  3.0 &   \phn3$\pm$3 & 1 \nl
\enddata
\tablenotetext{a}{NGC 681 is one of our sample galaxies and contained 
also in Table~1.}
\tablecomments{Col.~(1): galaxy name; 
cols.~(2)-(3): right ascension and declination (2000.0 coordinates) from RC3; 
col.~(4): morphological type from RC3; 
col.~(5): $B_T^0$ from RC3; 
col.~(6): recession velocity $V_{GSR}$ from RC3; 
col.~(7): optical angular diametar caluculated usuig $D_{25}$ of RC3; 
col.~(8): equivalent width of H$\alpha$+[N{\small II}].}
\tablerefs{(1)~Kennicutt 1983; (2)~Romanishin 1990.}
\end{deluxetable}

\clearpage
\begin{deluxetable}{ccccccc}
\scriptsize
\tablecolumns{7}
\tablewidth{0pt}
\tablenum{3}
\tablecaption{Observed EW(H$\alpha$) and Other Results}
\tablehead{
\colhead{Galaxy}  & \colhead{~~EW(H$\alpha$)} & \colhead{log$f_{{\rm H\alpha}}$} & 
\colhead{log$L_{{\rm H\alpha}}$} & \colhead{$r_{e,{\rm H\alpha }}$} & 
\colhead{$r_{e,{\rm H\alpha }}/r_{25}$}  & \colhead{$r_{e,{\rm H\alpha }}/r_{e,R}$} \\
\colhead{} & \colhead{~~(\AA )} &  \colhead{(erg~s$^{-1}$~cm$^{-2}$)}  & 
\colhead{(L$_{\odot}$)} & \colhead{(kpc)} & \colhead{} & \colhead{} \\
\colhead{(1)} &  \colhead{~~(2)} &  \colhead{(3)} & \colhead{(4)} & 
\colhead{(5)} & \colhead{(6)} & \colhead{(7)} \\ 
\cline{1-7}
\tablevspace{0.08in}
\multicolumn{7}{c}{the comparison galaxies}  }
\startdata
NGC  3055      &  ~~50$\pm$8.2 & $-$11.7 &  7.51 & 1.4 &0.21& 0.65 \nl
NGC  6207      &  ~~46$\pm$1.7 & $-$11.6 &  7.21 & 1.4 &0.24&0.96 \nl
NGC  6574      &  ~~36$\pm$0.8 & $-$11.8 &  7.96 & 1.9 &0.24&0.75 \nl
NGC  7217      &  ~~14$\pm$1.6 & $-$11.5 &  7.31 & 3.0 &0.31&1.21 \nl
NGC  7448      &  ~~44$\pm$0.6 & $-$11.6 &  7.97 & 3.5 &0.28&1.21 \nl
NGC  7743      &  ~~\phn3$\pm$1.3 & $-$12.7 &  6.56 & 0.4 &0.03&0.10 \nl
\cutinhead{the sample galaxies} 
NGC   ~681\tablenotemark{a}  &  ~~20$\pm$2.1 & $-$12.1 &  7.16  & 2.5 &0.28 & 1.05\nl
NGC  1022      &  ~~14$\pm$2.0 & $-$11.9 &  7.12  & 0.4 &0.06 &  0.17\nl
NGC  2782      &  ~~33$\pm$1.5 & $-$11.7 &  7.84  & 1.0 &0.06 & 0.30\nl
NGC  2993      & ~~114$\pm$1.6\phn & $-$11.6 &  7.91  & 0.7 &0.12 & 0.56\nl
NGC  3442      &  ~~69$\pm$1.9 & $-$12.1 &  7.10  & 0.7 &0.33 & 1.00\nl
NGC  3504      &  ~~51$\pm$1.3 & $-$11.2 &  7.66  & 0.6 &0.07 & 0.29 \nl
NGC  3611      &  ~~27$\pm$1.4 & $-$12.0 &  7.09  & 0.6 &0.09 & 0.50 \nl
NGC  3729      &  ~~31$\pm$3.2 & $-$11.6 &  7.15  & 1.5 &0.26 & 0.79 \nl
NGC  4045      &  ~~17$\pm$2.4 & $-$12.1 &  7.13  & 2.2 &0.21 & 0.90 \nl
NGC  4369      &  ~~34$\pm$2.9 & $-$11.7 &  7.02  & 0.5 &0.11 & 0.44 \nl
NGC  4384      &  ~~37$\pm$4.0 & $-$12.2 &  7.33  & 1.2 &0.18 & 0.58 \nl
NGC  5534      &  ~~45$\pm$3.0 & $-$11.8 &  7.73  & 1.8 &0.26 & 0.95\nl
NGC  5691      &  ~~40$\pm$2.0 & $-$11.7 &  7.60  & 1.2 &0.18 & 0.62\nl
NGC  5915      &  ~~70$\pm$3.2 & $-$11.6 &  7.93  & 1.5 &0.18 & 0.77\nl
NGC  7625      &  ~~40$\pm$4.0 & $-$11.8 &  7.47  & 0.9 &0.17 & 0.57\nl
\enddata
\tablenotetext{a}{NGC 681 is included for the comparison galaxies as well.}
\tablecomments{Col.~(1): galaxy name; 
col.~(2): observed equivalent width of H$\alpha$+[N{\small II}] and its error; 
col.~(3): estimated H$\alpha$+[N{\small II}] line flux. See text for details; 
col.~(4): estimated H$\alpha$+[N{\small II}] luminosity; 
col.~(5): half-light radius of H$\alpha$ emission in kpc. See text
for details; 
col.~(6): the ratio of the half-light radius of H$\alpha$ emission to
the optical radius, $r_{25}$, which is a half of $D_{25}$ in 
Table~1 and Table~2; 
col.~(7): the ratio of the half-light radius of H$\alpha$ emission to that of 
$R$ band light.}
\end{deluxetable}


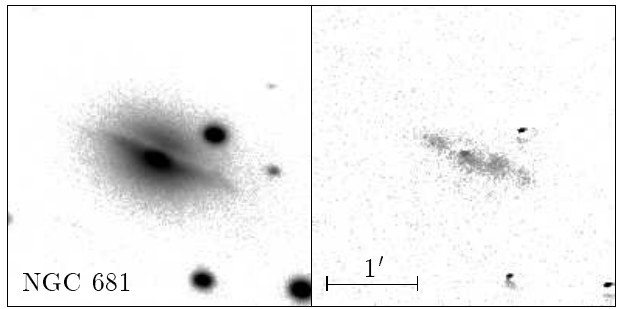
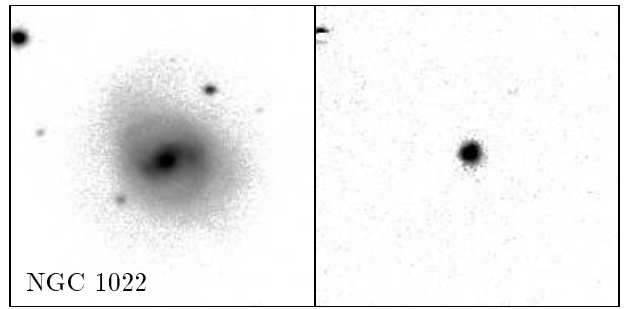
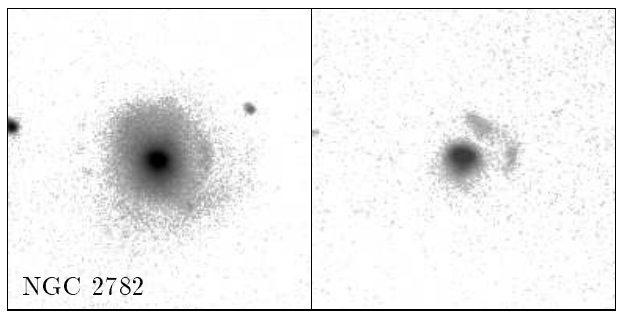
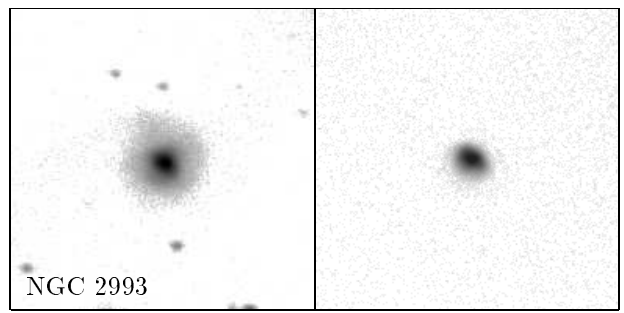
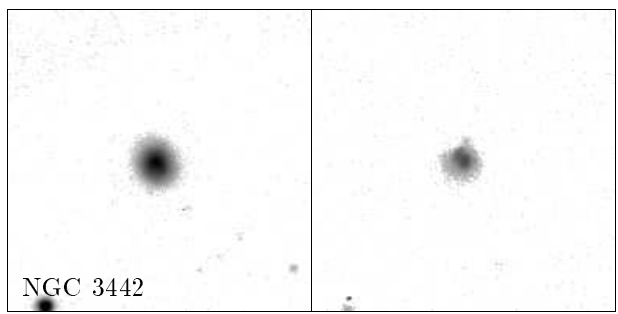
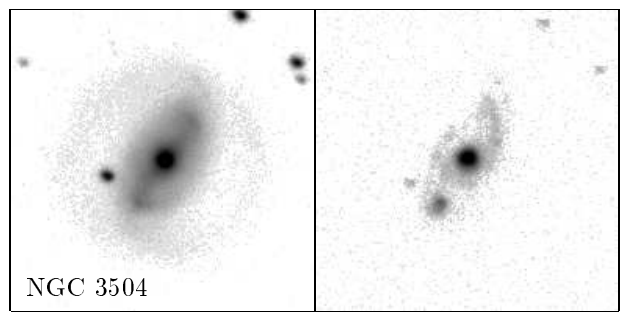
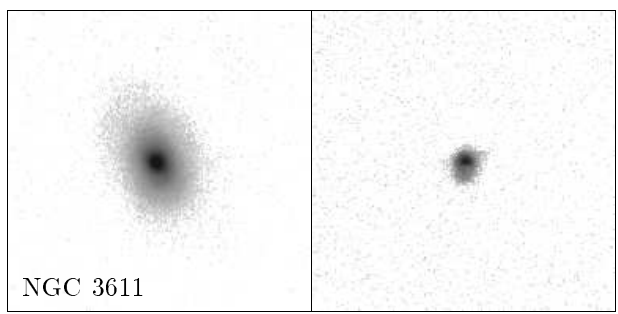
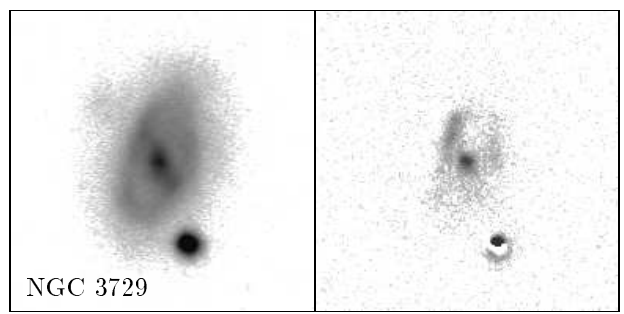
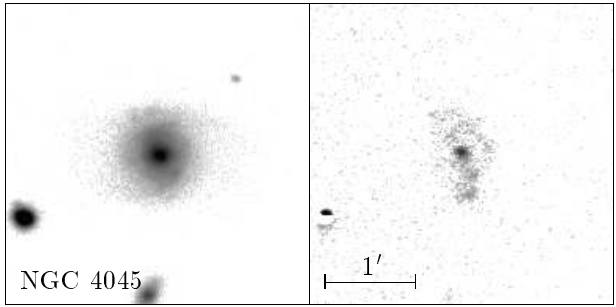
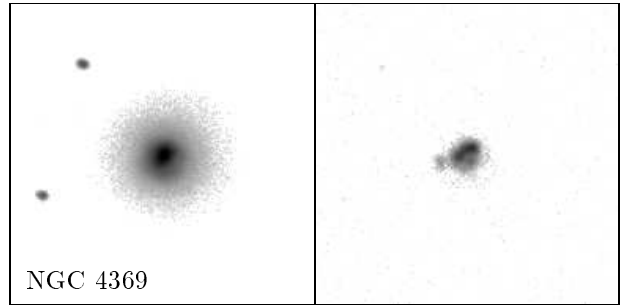
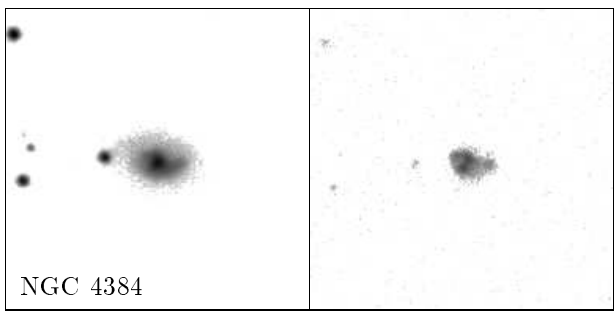
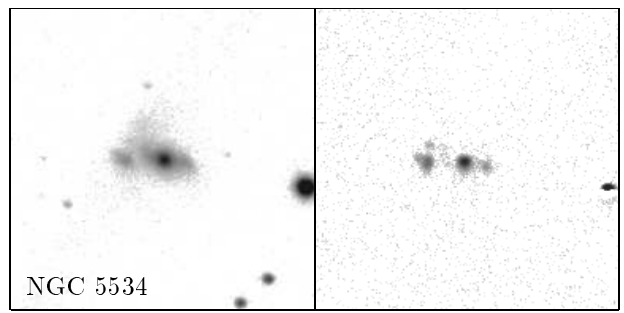
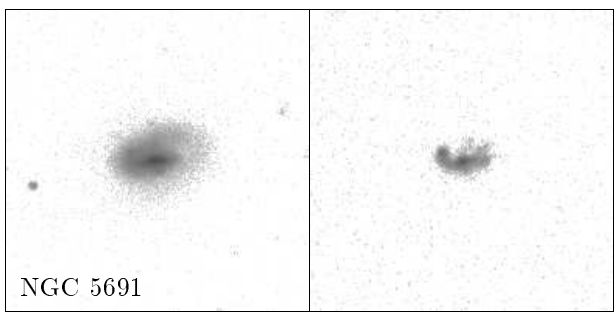
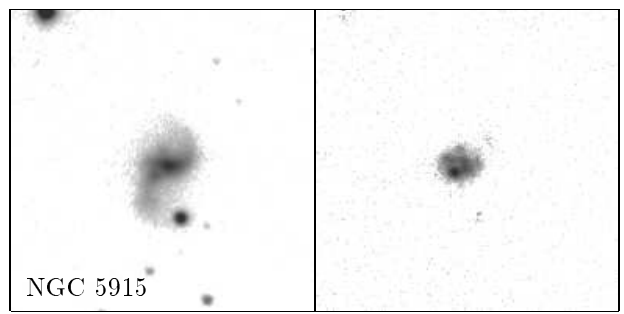
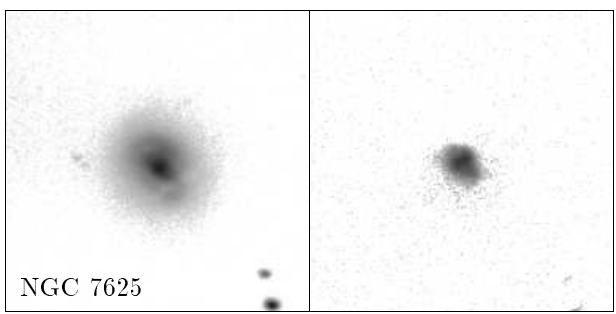
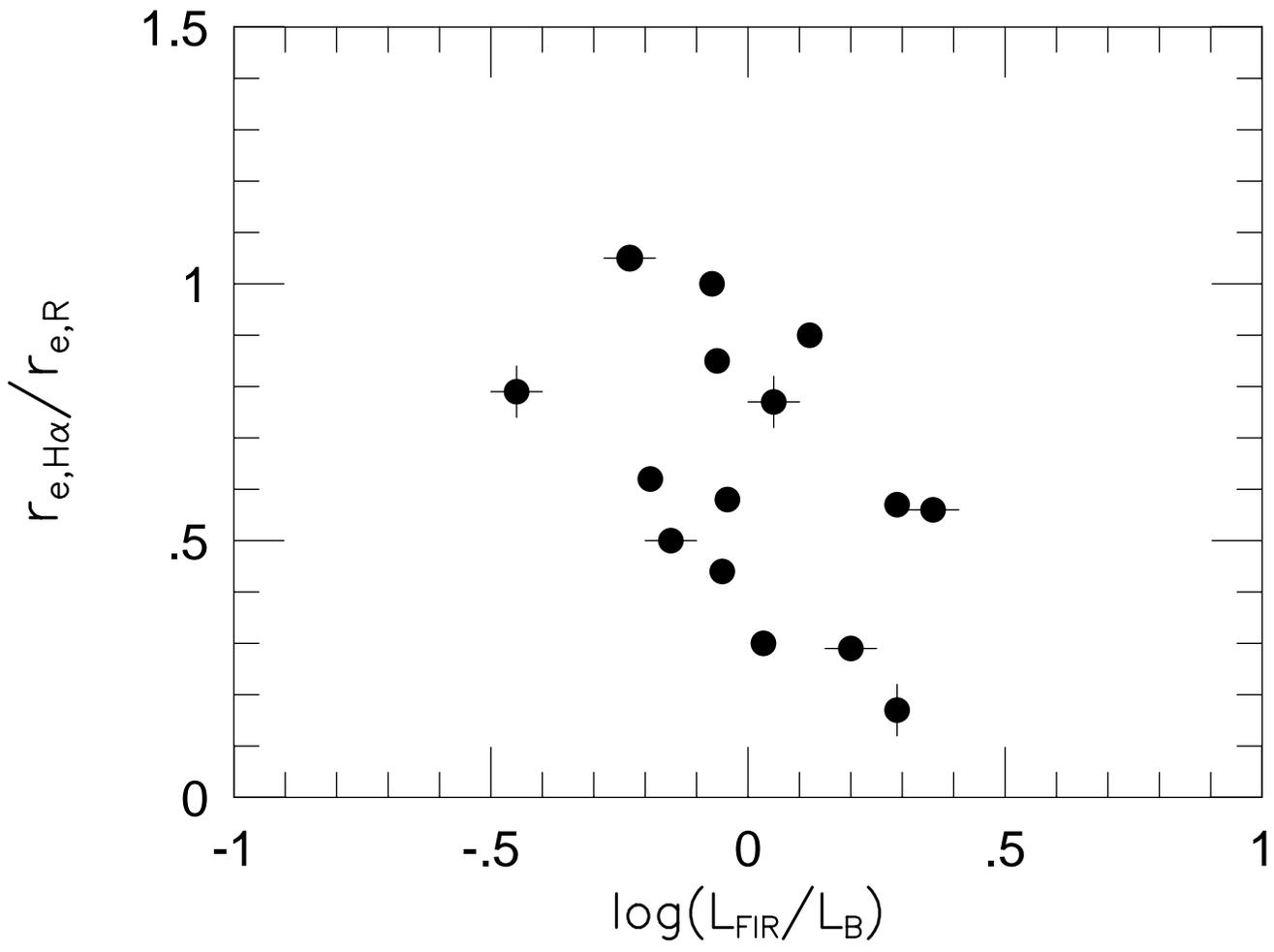
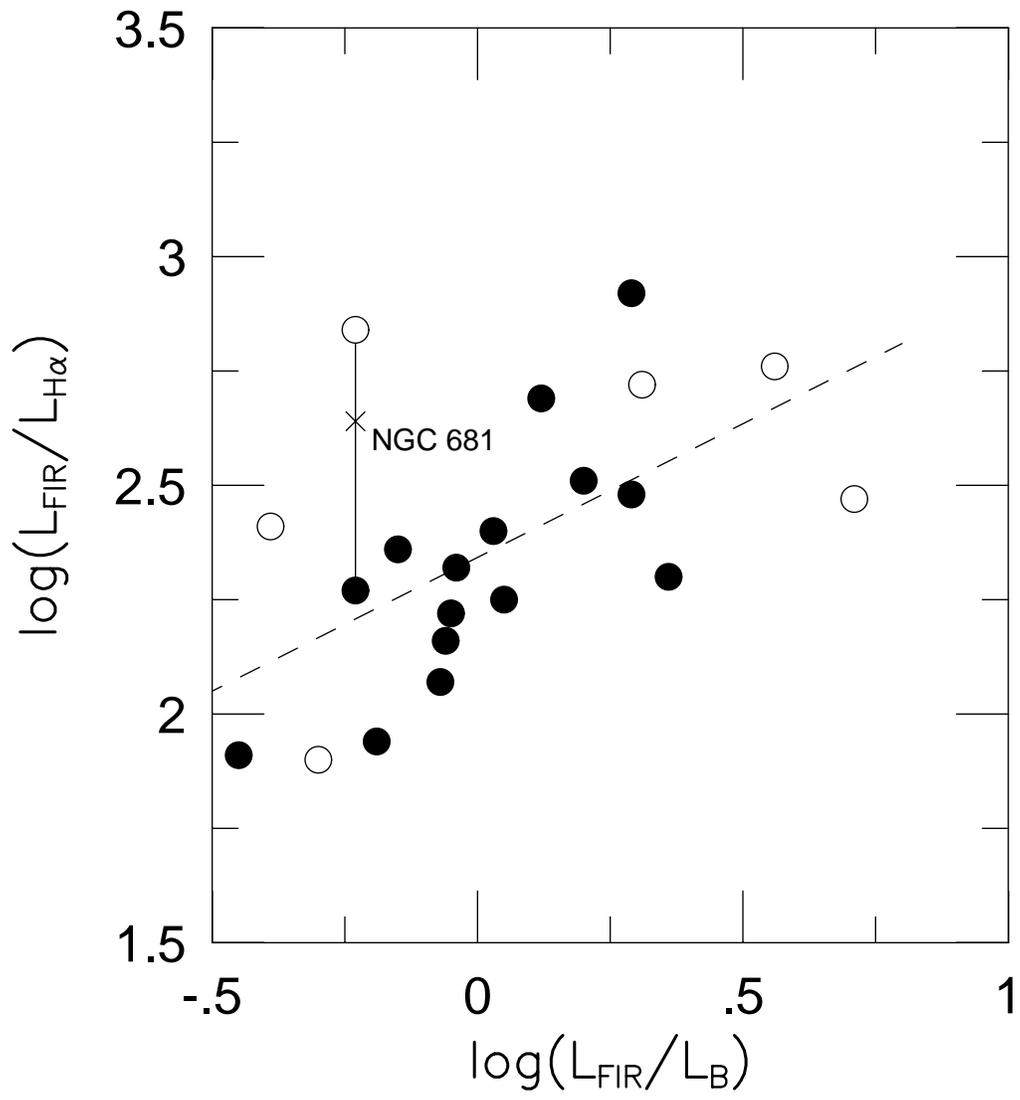
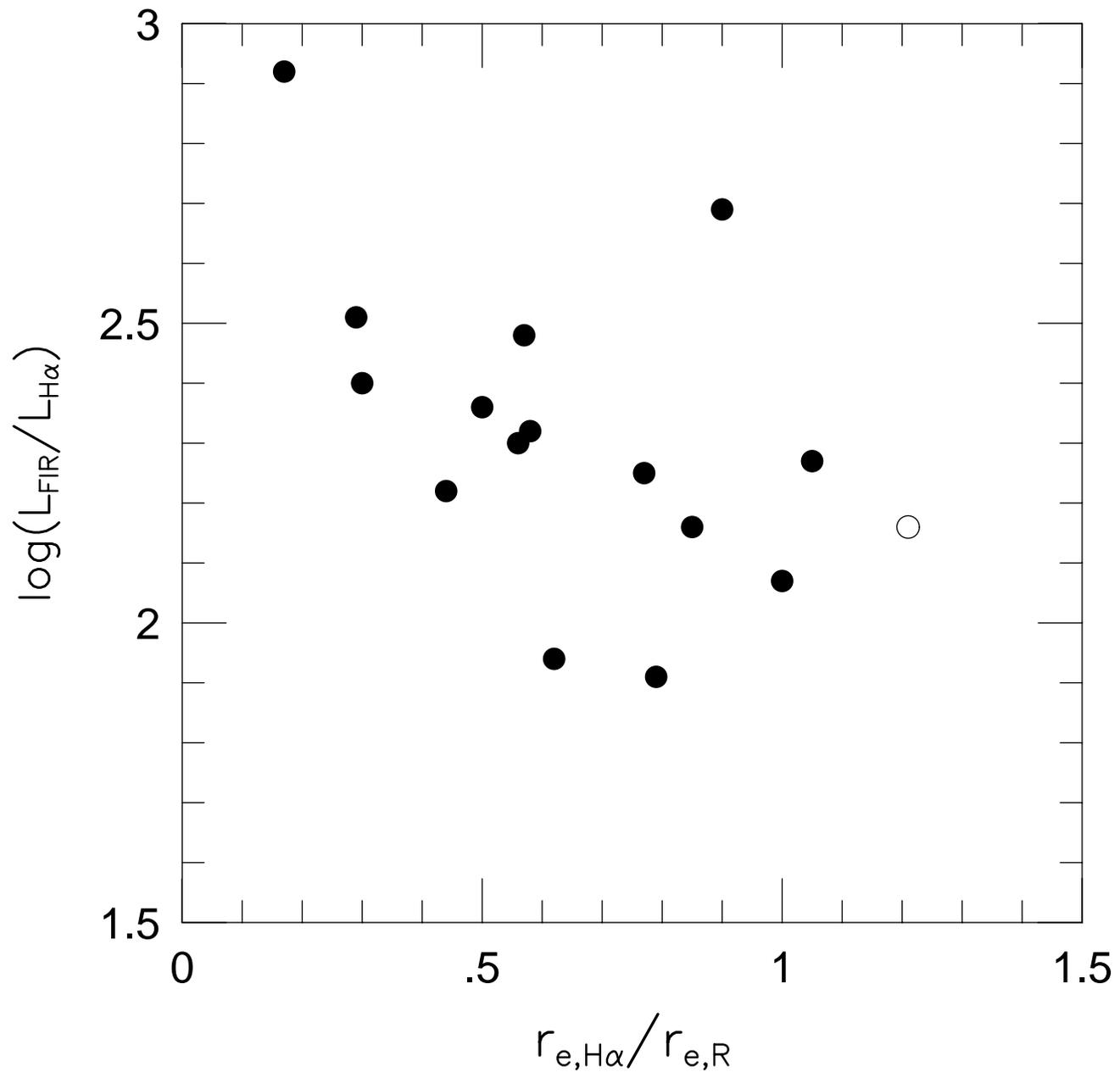
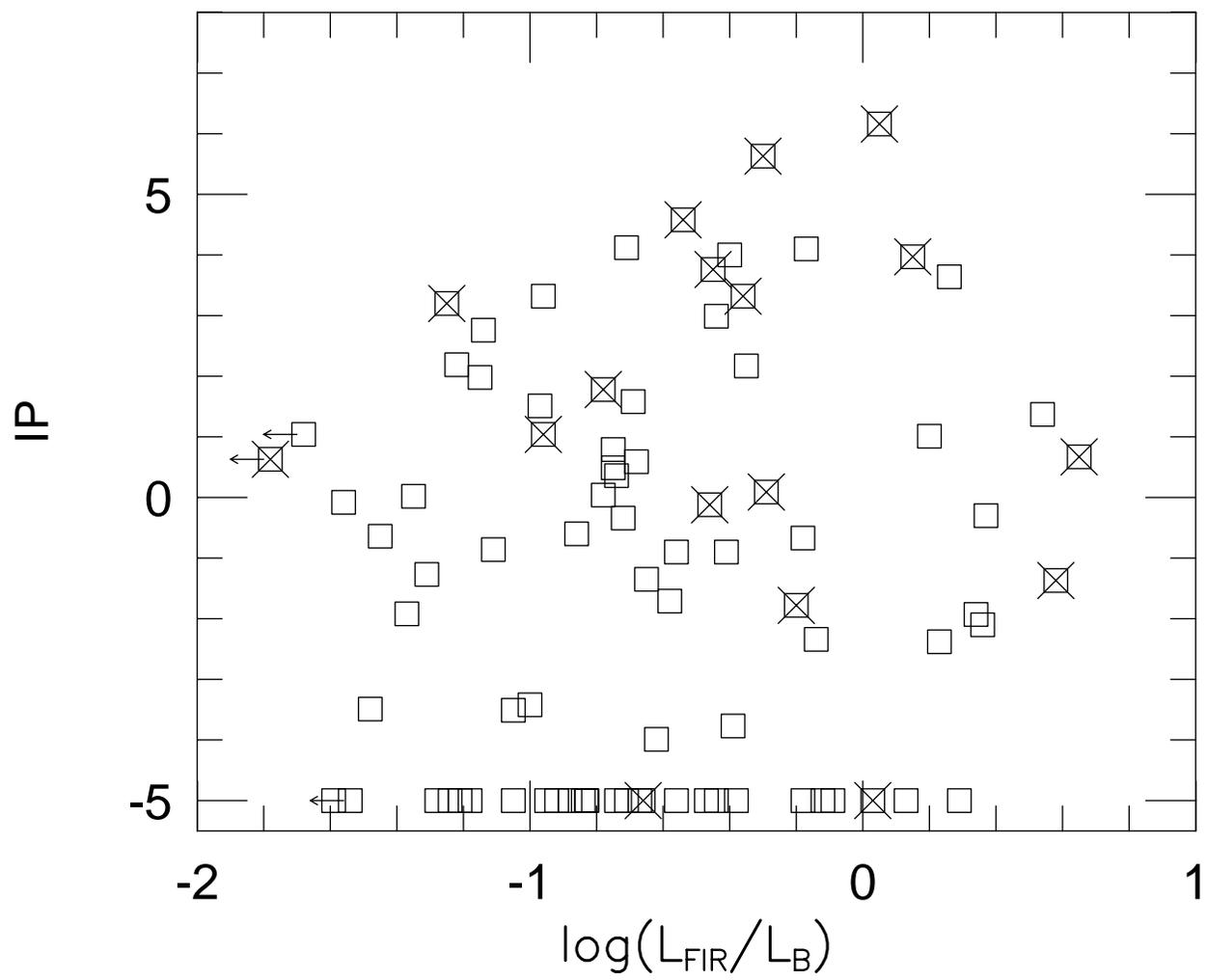
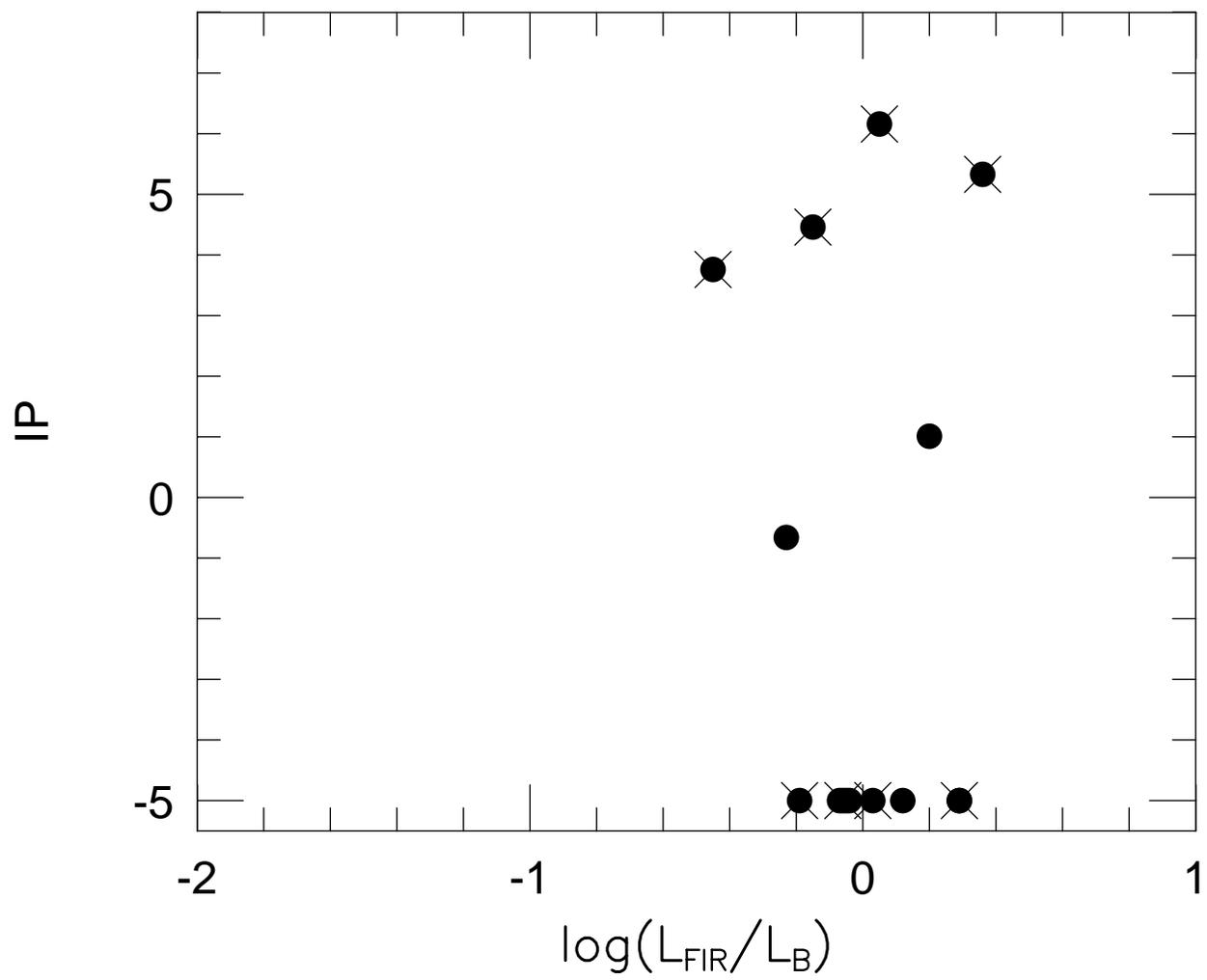
\end{document}